# A SURVEY ON ENERGY EFFICIENCY FOR WIRELESS MESH NETWORK

Sarra MAMECHAOUI [1], Fedoua DIDI[2] and Guy PUJOLLE [3]

[1,2] STIC Laboratory, Abou Bekr Belkaid University Tlemcen, Algeria
{sarra.mamechaoui, f_didi}@mail.univ-tlemcen.dz

[3] Pierre et Marie Curie University Paris 6, FRANCE
Guy.Pujolle@lip6.fr

## ABSTRACT

*Reducing $CO_2$ emissions is an important global environmental issue. Over the recent years, wireless and mobile communications have increasingly become popular with consumers. An increasingly popular type of wireless access is the so-called Wireless Mesh Networks (WMNs) that provide wireless connectivity through much cheaper and more flexible backhaul infrastructure compared with wired solutions. Wireless Mesh Network (WMN) is an emerging new technology which is being adopted as the wireless internetworking solution for the near future. Due to increased energy consumption in the information and communication technology (ICT) industries, and its consequent environmental effects, energy efficiency has become a key factor to evaluate the performance of a communication network.*

*This paper mainly focuses on classification layer of the largest existing approaches dedicated to energy conservation. It is also discussing the most interesting works on energy saving in WMNs networks.*

## 1. INTRODUCTION

Global energy consumption is currently one of the major concerns faced by governments worldwide, because of its significant environmental footprint and the eventual exhaustion. In a not-so-far future, the major traditional energy sources will be replaced by an alternative source using a sustainable energy source such as solar or wind power. Green networking has recently drawn a lot of attention, which comprises of a rethinking of the way networks are built and operated so that not only costs and performance are taken into account but also their energy consumption and carbon footprint.

The application of green networking to Wireless Mesh Networks (WMN) has seldom been reported in the literature. Wireless Mesh Network (WMN) [1] is an emerging new technology which is being adopted as the wireless internetworking solution for the near future. Characteristics of WMN such as rapid deployment and self configuration make WMN suitable for transient on-demand network deployment scenarios such as disaster recovery, hard-to-wire buildings, conventional networks and friendly terrains. Wireless Mesh Network is now being extensively used as a cost-effective means for coverage extension and backhaul relaying between IEEE 802.11 access points.

This solution aims to implement mobility in wireless mesh networks. Indeed, Mesh technology has captured the interest of academic research and industry, because of its ability to satisfy both the requirements of Internet Service Providers (ISPs) and wireless users. The Figure 1 show an example of a WLAN Mesh Network architecture [2] that consist of nodes called Mesh points mesh (Mesh Point MP) which only relay traffic but do not provide wireless coverage for mobile station (MSs). These MPs are using the services of wireless mesh network (WLAN Mesh) to communicate with other MPs in the network. They can act as an access point (Mesh Access Point MAP) or gateway (Mesh Point Portal: MPP) whose aim to integrate WMNs with various existing wireless networks such as cellular systems, wireless sensor networks, wireless-fidelity (Wi-Fi) [3] systems, worldwide inter-operability for microwave access (WiMAX) [4].

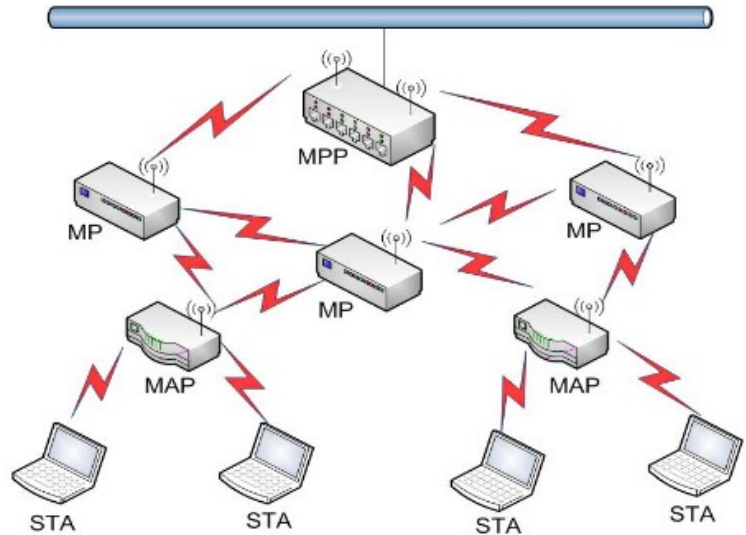

Figure 1: WLAN Mesh Networks architecture.

Due to increased energy consumption in the information and communication technology (ICT) industries, and its consequent environmental effects, energy efficiency has become a key factor to evaluate the performance of a communication network. In WMNs, the resources of Wireless Access Networks are, for long periods of time, underemployed, since only a few percentage of capacity of these devices is effectively used and these results in high energy waste [5]. This means that the energy consumption does not decrease when the traffic is low and that it would be possible to save large amounts of energy just by switching off unnecessary network elements.

In this context and in order to build green WMNs, it is important to design energy efficient planning and management strategies. This Paper addresses the incorporation of energy conservation at three layers lower of the protocol stack for wireless mesh networks. The remainder of this paper is organized as follows. Section 2 introduces the protocols that save energy in the network layer. Section 3 describes work dealing with energy efficient protocols in the MAC layer of wireless mesh networks, and power conserving protocols in the physical layer are addressed in section 4. Finally, section 5 summarizes and concludes the paper.

## 2. ENERGY CONSERVATION IN NETWORK LAYER

The network layer is responsible for network self-configuration and data routing. For configuring network topology, the network layer select an appropriate mode for a node and determines the most suitable neighbors with which to associate and form communication links.

The routing protocols enable a network to make dynamic adjustments to its conditions; these decisions do not have to be predetermined and static. Even though the routing protocols functions are plausible, they still face challenges in energy efficiency.

Routing protocols can be classified into three, namely: Table-driven, the Demand-driven and the Hybrid routing protocols. The table-driven routing protocols are also known as proactive routing protocol; each node maintains one or more tables which have routing information to all other nodes within the network. In this protocol, all nodes update each other on the same network and further update their tables to maintain a consistent and up-to-date view of the network. Demand-driven well known as reactive routing protocols creates routes only when desired by the source node [6]. When a node requires a route to its destination, it initiates a route discovery process within the network. The process is completed once a route is found or all possible route permutations had been examined. The hybrid routing protocols combines both the table-driven and the Demand-driven routing protocols to transport the packets from the source to the destination. It takes both the advantages of table-driven and on demand-driven routing protocols. One of the popular Hybrids routing protocol in the WMN is Hybrid Wireless Mesh Protocol (HWMP) [7]. This is a default routing protocol for IEEE 802.11s, which is consists of two features, namely: it is table-driven; and also an on-demand protocol. This protocol is based on the protocol called RM-OADV (Radio Metric-Ad Hoc on-Demand Vector) routing protocol, which is the extension of AODV. It uses the same route discovery mechanism as that used by AODV and Dynamic Source routing (DSR). There is therefore a need for a routing protocol that will be energy efficient and also scalable.

Different technique of power saving for multi-hop ad hoc wireless networks that reduces energy consumption have been proposed without diminishing of the capacity or connectivity of the network. As WMNs share many common features with ad hoc networks. Thus, the approaches of energy conservation for ad hoc networks can usually be applied to WMNs. We find essentially the following approaches:

## 2.1. The CDS (Connected Dominating Set)

The CDS [8] use information of neighborhood or topology to determine the set of nodes which form a connected dominating set (CDS) for the network where all nodes are either a member of the CDS or a direct neighbor of at least one member. Nodes in the CDS are considered the pivotal routing and remain active all the time in order to maintain global connectivity. All other nodes can choose to sleep if necessary. Figure 2, show the set {A, B, C, D, E, F} are the member of the CDS and remain active all the time to deliver the flow and maintain the connectivity to their neighbor. All other nodes {1 to 18} can choose to sleep if there is no data to transmit.

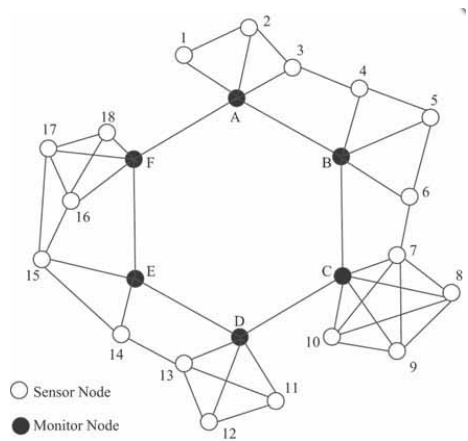

Figure 2: An example of Connected Dominating Set (CDS).

## 2.2. The SPAN

In SPAN [9], each node in the network using Span protocol can make periodic, local decisions on the desirability to sleep or stay awake as a coordinator and take part in the transmission backbone topology. Each node decides to be a coordinator or not. The transition between the two states is made based on probabilities. Fairness is ensured by the node to a higher energy the most likely to be a coordinator. The other criterion used in selecting coordinators is the value that a node adds to the overall network connectivity. A node connecting more nodes will be more likely to be chosen as coordinator. The notion of randomness is used to prevent multiple simultaneous coordinators. Span adaptively elects "coordinators" from all nodes in the network. Span coordinators stay awake continuously and perform multi-hop packet routing within the ad hoc network, while other nodes remain in power-saving mode and periodically check if they should wake up and become a coordinator. The main disadvantage of this protocol is that it is inherently dependent on node density for energy savings [12].

## 2.3. The GAF

GAF [10] is another technique that uses knowledge of the geographic positions of nodes to select coordinators. The geographic positions of nodes are used to divide the complete topology in areas of fixed size (fixed geographic area). Areas are created such that any two nodes in any two adjacent zones can communicate. The size of the zone is thus dictated by the radio range of the nodes which is supposed to be fixed. Only one node in each zone must be awake and may be the coordinator. So, by exploiting the knowledge of the geographical positions GAF simplifies the selection process coordinator. The main disadvantage of this protocol is requiring that all nodes in the network know their geographic positions.

## 2.4. EMM-DSR protocol

In [11], authors suggest a new mechanism that allows making a trade-off between energy efficiency and the shortness of a selected path for forwarding data packets. In other word, this mechanism tries to minimize the energy consumption and, at the same time, maintain a good end-to-end delay and throughput performance. Their solution consists in extending the Max-Min algorithm to support the cited trade-off. Thus, they incorporate this extension, among other options, to the existing on-demand dynamic source routing protocol (DSR), and the resultant version takes the name of EMM-DSR (Extended Max-Min DSR).

## 2.5. The Minimum-energy Routing

Minimum-energy routing [12] saves power by choosing paths through a multi-hop ad hoc network that minimize the total transmit energy. This approach has been extended by Chang and Tassiulas [13] to maximize overall network lifetime by distributing energy consumption fairly. In this protocol, nodes adjust their transmission power levels and select routes to optimize performance.

## 2.6. Power-Aware Routing

In [14] a power-aware routing algorithm is presented for wireless networks with renewable energy sources. The proposed algorithm is shown to be asymptotically optimal. No information is assumed regarding the arrival process and it is assumed that the node has full knowledge of the energy it will receive until the next renewal point by looking at previous data. The proposed routing algorithm uses a composite cost metric that includes power for transmission and reception, replenishment rate, and residual energy. The work also includes non-uniform energy replenishment rates and introduces a battery energy threshold scheme to decrease overhead.

## 2.7. The Pulse

The Pulse protocol [15] design is centered around a flood refer to as a pulse, which is periodically sent at a fixed pulse interval. This pulse flood originates from infrastructure access nodes (pulse sources) and propagates through the entire component of the network. This rhythmic pulse serves two functions simultaneously. First, it serves as the primary routing mechanism by periodically updating each node in the networks route to the nearest pulse source. Each node tracks the best route to the pulse source by remembering only the node from which it received a flood packet with the lowest metric. The propagation of the flood forms a loop free routing tree rooted at the pulse source. Second, if a node needs to send and receive packets, it responds to the flood with a reservation packet. This reservation packet is sent up the tree to the pulse source. The reservation packet contains the address of the node making the reservation, and is used to setup reverse routes at all nodes on the path between the pulse source and the sending node. A node that does not send or forward a reservation packet will have no packet forwarding responsibilities until the next pulse occurs, it may place its radio in sleep mode until the next pulse period begins. This node deactivation is what allows the Pulse protocol to conserve power. Figure 3, shows the Pulse protocol in an example network. Every node in the example network has a route towards the pulse source as indicated by the grey arrows. Both nodes A and B are actively communicating, and have each sent a reservation packet up the tree to the pulse source. The reservation packets have setup reverse routes as indicated by the black bi-directional arrows. Nodes that have forwarded a reservation stay "on" to forward data which are with the black. The rest of the nodes in the network may turn "off" until the next pulse.

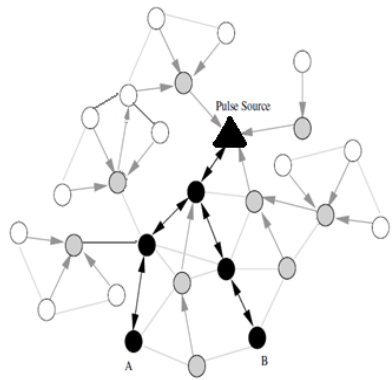

Figure 3: The Pulse protocol in an example network.

This protocol [30] has significant disadvantages in the routing information that is essentially the problem of flooding packets and the problem of Overlap delay or it can result in a significant consumption of energy.

## 2.8. The green-clustering

The green-clustering algorithm [16] was developed to enable the central controller in a WLAN to make certain decisions to power on and off portions of it based on certain pre-defined criteria like deployment, location of access points, and locally derived information. Hence, this algorithm was designed with respect to the centralized network infrastructure and its devices. The basic idea was that if access points are close enough in the cluster, a single one would fulfill the needs of users, even those in the vicinity of other APs in the same cluster. This clustering concept was designed with respect to large organizations with high-density WLANs because access points are placed very close to each other to provide overlapping coverage and high bandwidth.

### 2.9. CaDet (clustering and decision-tree-based)

The clustering and decision-tree-based method for reducing power consumption was also approached by algorithms used on location estimation systems [17], where users are located based on the wireless signals employing a wireless device with minimum computational resources and power requirements. Using a combination of data-mining techniques and analysis of real wireless data by probabilistic location estimation and the multiple-decisions tree-based approach, some very good results can be found. These results would help to identify the most utilized access points in the network, which in turn is an easy way to predict client locations. These decisions would allow a minimum of APs to be used, thus reducing the power consumption and wake-up time of the client.

### 2.10. EAR (Energy Aware Routing)

The main idea of energy aware routing (EAR) [18] is to take energy consumption of equipment into account when doing routing and traffic-engineering decisions. In this respect, energy consumption or energy cost can be used stand-alone or in combination with other objectives and constraints like QoS or availability characteristics. EAR concept aims to switch-off completely components which certainly help to reduce the energy consumption of a network. Therefore they propose to extend the concept towards an energy profile aware routing (EPAR) and to include energy profiles into dimensioning, routing and traffic engineering decisions.

### 2.11. ETR (Throughput-aware Routing)

In [19], authors proposed a novel routing solution for 802.11 based-WMNs to provide flows with throughput guarantees while optimizing the energy consumption. The key contributions of this paper are: First, they proposed a novel way to represent the capacity region of a set of stations sharing the wireless medium, which they refer to as linearized capacity region, and compute the corresponding parameters that define this region for the case of 802.11. The linearized capacity region is devised with the aim of aiding the design of optimal and efficient networking algorithms. Second, based on the information provided by the linearized capacity region, they design a novel optimal routing algorithm for wireless mesh networks that admits as many users as possible to the network while providing them with the desired throughput guarantees. Finally, they design an extension to their routing algorithm that, in addition to providing throughput guarantees, minimizes the total energy consumed by the mesh network. This is achieved by switching off as many mesh routers as possible.

### 2.12. Green Framework

In [20], authors proposed a new framework for energy management in TDMA based WMNs to support energy efficient communications. The proposed green framework provided the WMN administrator with a parameterized objective function to choose to yield the desired trade-off between the achieved network throughput and energy consumption. First, they proposed an Optimal Green Routing and Link Scheduling, called O-GRLS, which aims at finding the optimal trade-off. The problem was formulated as an integer linear program (ILP). As this problem is known to be NP-hard, they then proposed a simple yet efficient algorithm based on Ant Colony, called Ant Colony Green Routing and Link Scheduling (AC-GRLS) to solve the formulated ILP problem. In this context, the cases of O-GRLS, a Beam Search heuristic, and Shortest Path (SP) routing strategy, are used to develop baselines to which the AC-GRLS improvements are compared. Through extensive simulations, they showed that their framework can achieve significant gains in terms of energy consumption as well as achieved network throughput, compared to the Shortest Path (SP) routing, as well as the Beam Search heuristic. Specifically, in small sized networks, for grid and random networks, the proposed energy management approaches can save, respectively, 29% and 20% of the energy cost, while achieving the same performance as SP. However, if the network consumes the same energy as SP, the achieved throughput can be enhanced by up to 30%. These gains are maintained in large-sized networks. Indeed, the simulation showed that the energy saving is about 20%, while the achievable throughput improvement is about 20% for both grid and random network topologies. In addition, they showed that AC-GRLS converges to the optimal solution in small sized WMNs and has low computation time in large-sized ones, which makes it a feasible and efficient solution for green WMNs and energy efficient management.

### 2.13. CBRP (Cluster Based Routing Protocol)

CBRP (Cluster Based Routing Protocol) [21] is an on-demand routing protocol, where the nodes in the network are divided into clusters. This protocol used clustering structure for routing. Each cluster has a cluster head (CH) as coordinator. Each cluster head acts as a temporary base station within cluster and communicates with other cluster heads. CBRP is designed to be used in Wireless sensor network and mobile ad hoc network. The protocol divides the nodes of the ad hoc network into a number of overlapping or disjoint 2-hop diameter clusters in a distributed manner. Each cluster chooses a head to retain cluster membership information. There are four possible states for the node: Normal, Isolated, Cluster head (CH) and Gateway. Initially all nodes are in the state of Isolated. Each node maintains the neighbor table where in the information about the other neighbors nodes is stored; cluster heads have another table (cluster heads neighbor) where include the information about the other neighbor cluster heads is stored. The protocol efficiently minimizes the flooding traffic during route discovery and speeds up this process as well or it may result in significant energy savings.

**Summary**

The power saving protocols of Network Layer presented in this section is summarized in the table below.

| Protocol | Network | Topology | Contribution |
| --- | --- | --- | --- |
| CDS [8] | Ad hoc | Flat | - Use information of neighborhood to determine the set of nodes which form a CDS which are considers as the pivotal routing and remain active all the time. |
| SPAN [9] | Ad hoc | Grid | - Local decisions on the desirability to sleep or stay awake as a coordinator and take part in the transmission. |
| GAF [10] | Ad hoc | Grid | - Uses knowledge of the geographic positions of nodes to select coordinators. |
| EMM-DSR [11] | Ad hoc | Flat | - The selection of the shortness path for forwarding data packets. |
| Minimum-energy Routing [12] | Ad hoc | Flat | - Node adjusts it transmission power levels and select routes to optimize performance. |

| Power-Aware Routing [14] | Ad hoc | Flat | - Routing algorithm uses a composite cost metric that includes power for transmission and reception, replenishment rate, and residual energy. |
|---|---|---|---|
| Pulse [15] | multi-hop wireless infrastructure | Tree | - Flood refers to as a pulse, which is periodically sent at a fixed pulse interval to tracks the best route to the pulse source. |
| Green-clustering [16] | WLAN infrastructure | Clustered | - Central control to make certain decisions to power on and off portions of it based on certain pre-defined criteria. |
| CaDet [17] | WLAN infrastructure | Clustered | - Clustering and decision-tree-based. |
| EAR [18] | All networks | Flat | - Take energy consumption of equipment into account when doing routing and traffic-engineering decisions. |
| ETR [19] | Wireless Mesh Networks | Flat | - Routing algorithm that uses as few nodes as possible, this allows switching off the unused nodes. |
| Green Framework [20] | Wireless Mesh Networks | Flat | - Optimal Green Routing and Link Scheduling.<br>- Ant Colony Green Routing and Link Scheduling. |
| CBRP [21] | Wireless Mesh Networks | Clustered | Minimizes the flooding traffic during route discovery. |

In terms of routing, we think the approach of clustering is the most interesting to conserve energy consumption. The selection of the cluster head in each cluster on the basis of energy level in wireless mesh network can reduce the rate of energy consumption by scheduling activities in the cluster of mesh users but not for mesh routers. The Pulse protocol is designed for multi-hop wireless infrastructure access and an extensive set of simulations had demonstrated that this protocol is effective at both routing and conserving energy.

## 3. ENERGY CONSERVATION IN DATA LINK LAYER

In the previous section we showed approaches that conserve energy in Network Layer. We now discuss some of the key issues relating to the implementation of power saving at MAC Layer sub-layer of Data Link Layer which provides a fair mechanism to share access to medium among other nodes. The MAC plays a key role in the maximization of node's energy efficiency.

We first briefly review power saving in conventional IEEE 802.11and discussion of recent work that has started to address them.

### 3.1 PSM (Power Saving Mode) IEEE 802.11 Protocol

The 802.11 standard defines a Power-Saving Mode (PSM) [22], aimed at reducing the energy consumption of mobile devices. The objective of the 802.11 PSM is to let the wireless interface of a mobile host in the active mode only for the time necessary to exchange data, and turn it in sleep mode whenever it becomes idle. PSM defines two diverse power management modes a mobile device can operate in one of them: active mode and Power Saving Mode (PSM). In active mode, a mobile device is fully powered and is able to exchange frames at any time. While in Power Saving mode, a mobile device is permitted to be in one of two different power states, either in awake state or doze state. An access point (AP) in a wireless network observes the mode of each mobile device. A mobile device must first inform its access point (AP) about changing its power management mode using Power Management bits within the Frame Control field of the frame used as a power saving request. A mobile device should not enter PSM before it receives an acknowledgement from the access point (AP). During the association procedure, a mobile device informs the access point AP about its listen interval, which is used to indicate a period of time for which a mobile device in PSM may choose to sleep. Its aim is to reduce energy consumption in conventional IEEE 802.11 standard.

A mobile device running PSM can go to sleep which incurs power consumption of only around 50mW. However, when a PSM device is sleeping, it cannot transmit or receive any packets; hence, PSM clients conserve energy at the cost of larger packet delivery delay.

### 3.2. APSD (Automatic Power Save Delivery)

Various enhancements to the above scheme have been included in some of the follow-on standards. For example, IEEE 802.11e defines Automatic Power Save Delivery (APSD) [23]. This includes both contention-based operation (referred to as EDCA) and a polling-based option (called HCCA). In the latter case, the access point (AP) functions as a hybrid coordinator (HC), and defines periodic service intervals that allow the synchronous delivery of traffic using Scheduled Automatic Power Save Delivery (S-APSD). In the former case the unscheduled APSD (U-APSD) mechanism permits the station to initiate communication activity by transmitting trigger frames on the uplink in EDCA contention mode. These mechanisms provide for improved flexibility and power saving compared to the original procedures.

### 3.3. PSMP (Power Save Multi-Poll)

IEEE 802.11n has introduced further enhancements to the U-APSD and S-APSD protocols, referred to as power save multi-poll (PSMP) [24]. As in its predecessors, there are scheduled (i.e., S-PSMP) and unscheduled (i.e., U-PSMP) versions. S-PSMP provides tighter control over the AP/station timeline by having the AP define a PSMP sequence that includes scheduled times for downlink and uplink transmissions. The ability to do this allows (non-AP) stations to remain in Doze mode during the times when other stations are scheduled to be using the channel and reduces AP/station interaction overheads. U-PSMP is similar to U-APSD in that it supports both triggered and delivery enabled modes.

### 3.4. NAV (Network Allocation Vector) and NAM (Network Allocation Map)

The access point (AP) uses network allocation vector blocking to prevent channel access to the AP when it is in Doze mode [25]. In conventional IEEE 802.11, a NAV is used at each station to implement a virtual carrier sense mechanism and to block stations from transmitting in cases where the channel has been reserved for some other purpose.

The IEEE 802.11 NAV mechanism is generalized, and a power saving AP includes a network allocation map (NAM) in its beacon broadcasts [26]. The NAM specifies periods of time within the super frame when the AP is unavailable, and during these periods the AP is assumed to be inactive and conserving power.

### 3.5. PAMAS (Power-aware Multi Access Protocol with Signalling)

The PAMAS [27] power-saving medium access protocol proposed to turning off a node's radio when it is overhearing a packet not addressed to it. It is a combination of original MAC protocol, and using a separate channel for a busy signal. By using the busy signal, the terminals are able to determine when and how long they should turn off their radio interfaces. In this protocol, if a node has no packet to transmit, then it should turn off its radio interface if one of its neighboring nodes begins transmitting. Similarly, if at least one neighboring node transmits and another receives, the node should also turn off power because it cannot transmit or receive packets (even if its transmission queue is not empty). This approach uses a separate channel for a busy signal. Each node listens to channel to see when it becomes free to transmit (idle-listening). So it leads to important energy consumption.

### 3.6. SOFA (A Sleep-Optimal Fair-Attention scheduler)

In SOFA [28] authors proposed a downlink traffic scheduler on the AP of a WLAN, called SOFA, which help its PSM clients to save energy by allowing them to sleep more, hence to increase battery lives. If a client has buffered packets at the AP in a beacon period, and that client decides to receive it, it has to remain awake from the beginning of the beacon period till the last packet scheduled for it in the beacon period is delivered. Therefore a large portion of energy wastage (for the client) comes from the AP transmitting other clients' a packets before it finishes transmitting the client's last packet to it. SOFA manages to reduce such energy wastage and maximizes the total sleep time of all clients. SOFA favors clients with smaller attention requests by allowing them to spend less energy to get one unit of attention, while still helping other clients with larger attention requests to sleep more compared with other popular scheduling policies like round-robin and FCFS.

### 3.7. PRCW (Physical Rate and Contention Window based admission control)

In [29], the paper present a class based admission control algorithm for 802.11e based wireless local area networks. The strengths of this admission control is dynamicity and flexibility of the algorithm, which adapts to the situation of the BSS, like global load, number of best effort AC, and position of QSTA by report of QAP, thing that have never been taken together, but each solutions have used a point of sight separately. Thus it achieves higher throughput than other admission control for 802.11e. The idea which consists of changing parameters [AIFSN, CWmin and CWmax] of best effort flows, for decreasing collisions, is used, so they think that it is an efficient way to protect QoS flows from best effort flows, and to allow reducing the number of collisions and power consumption. So they increase AIFSN [best effort flow], CWmin and CWmax only at 70% of load of network, to prevent starvations of best effort flows, and so they increase rate utilization of channel. They also use the current rate transmission of QSTAs, according to their positions, instead of the minimum rate transmission used by standard 802.11e, for calculate the load of network and derived the $TXOP_i$ necessary for all the stations, with i=1 to number of active stations. The 802.11e standard starves the low priority traffic in case of high load, and leads to higher collision rates, and did not make a good estimate of weight of queues, so there is an unbalance enters the flows with high priorities.

## 3.8. LEACH (Low-Energy Adaptive clustering Hierarchy)

LEACH is scheduled MAC protocol with clustered topology [30]. It is considered the first hierarchical routing protocol. It is also one of the hierarchical routing algorithms for WSN popular, it combines the efficiency of energy consumption and quality of media access, and based on the division into clusters, view to enable the use of the concept of data aggregation for better performance in terms of lifetime.

The communication architecture of LEACH is, similarly to the cellular network to form cells based on the signal amplitude, and using the cell headers to node routers as well. These cells are called clusters, as heads: Cluster-heads (CH). The cluster leaders are chosen randomly according to a specific algorithm election based on a likelihood function that takes into account various criteria such as the available energy of nodes. As Figure 4 shows, the nodes are responsible for collecting data, send them to their CH aggregate and transmit them, in turn, results aggregation node well as a communication unicast (one hop). CHs are responsible for carrying out the functions most expensive energy, namely communication with the node sinks which is assumed far, and all data processing (aggregation, fusion and data) to reduce the amount the transmitted data. This device saves energy since the transmissions are provided solely by CH rather than all nodes in the network. Therefore, LEACH achieves a significant reduction in energy dissipation.

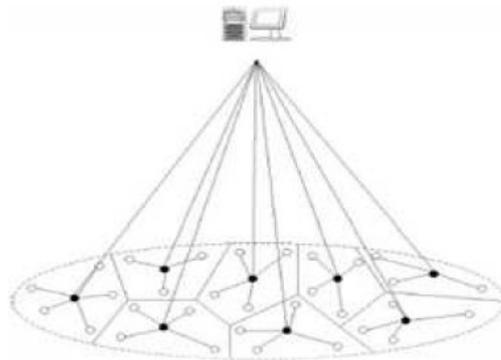

Figure 4: The architecture of Communications for LEACH protocol.

## 3.9. S-MAC

S-MAC [31] is an energy efficient MAC protocol for wireless sensor networks to reduce energy consumption, while supporting good scalability and collision avoidance. S-MAC adopts an effective mechanism to solve the energy wasting problems, that is periodical listening and sleeping. When a node is idle, it is more likely to be asleep instead of continuously listening to the channel. S-MAC reduces the listen time by letting the node go into periodic sleep mode. The main goal of S-MAC is to reduce power consumption including three major components: situation wake up and sleep is the periodic i.e. periodic sleep and listen, this protocol avoid the collision and overhearing meaning that in this protocol, nodes go to sleep after they hear an RTS or CTS packet and the duration field in each transmitted packet indicates how long the remaining transmission will be and communication between senders is the message passing that is shown in figure 5.

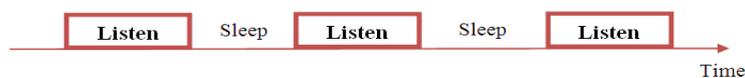

Figure 5: Periodic Listen and Sleep

## 3.10. Virtualization of NICs

In [32], one of the possible ways to save energy is to shutdown some of the APs, which serve a specific area of interest, if the number of the served stations are decreased in off-peak hours (e.g. at night, weekend, holidays). This could lead to the fact that some locations in the interested area are not covered, and stations located within these locations will not have connections. These can get connectivity by using Network coverage extension/ Relaying capability. In addition to the energy saved by shutting down some APs, NIC virtualization plays a role in reducing energy consumption.

A scenario represented in Figure 6 and Figure 7, in which a specific area are covered by a number of APs. This number is reduced by shutting down some of them if the number of the served stations is decreased in off-peak hours. In Figure 6, the whole area is served by six powered-on APs. Each circle represents an AP's coverage area. The symbols inside these cells represent stations. Stations with the same shape and colour are served by an AP. While in Fig. 3, only two APs are powered-on to reduce the consumed energy.

The number of APs is only reduced since the number of currently jointed stations is smaller than those in Figure 6. It's clearly shown in Figure 7 that the remaining powered-on APs do not cover the whole area. Thus, stations located outside (surrounded by dotted squares. These are equipped with virtualized NICs so that a station uses one of its virtual interfaces to create a network with other stations while keeping its connection to the serving AP using another virtual interface) the coverage range of the APs can get access to the Internet via stations (work as relays) connected to any served AP.

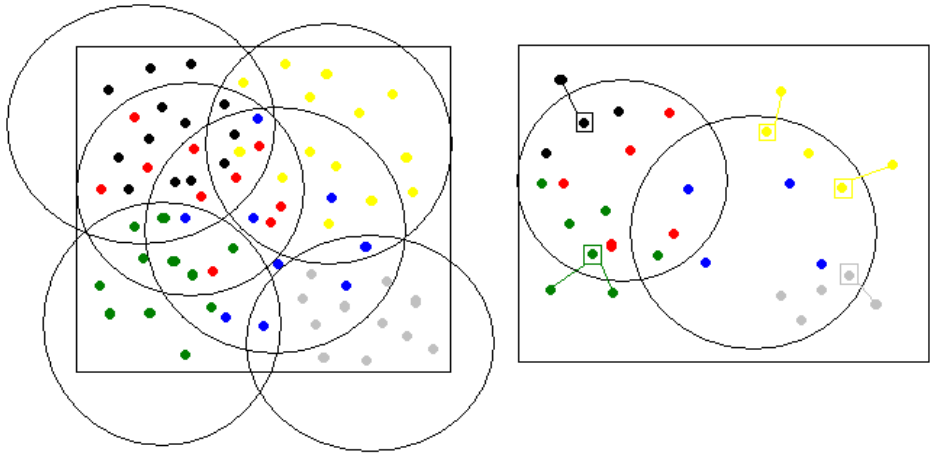

Figure 6: Stations served with 6 powered-on APs in a specific covered area.
Figure 7: Stations served with 2 powered-on APs in a specific covered area, some stations are equipped with virtual NICs to act also as APs and serve the stations existing in uncovered locations.

## 3.11. PEM (Power-Efficient MAC Protocol)

In [33] authors developed an efficient MAC protocol, PEM, to enhance the network utilization and reduce energy consumption. In PEM, stations can estimate their distances from the transmitter and obtain the interference relations among transmission pairs through three way-way handshaking (ATIM/ATIM-ACK/IIM). According to the interference relation, a scheduling algorithm is proposed for stations to schedule their transmissions. The scheduling algorithm can select interference-free transmission pairs as many as possible to increase the spatial reuse.

Thereafter, a station intending to transmit can adjust its transmission power to transmit at its scheduled time. In PEM, all stations know when to wake up and when they can enter doze state. Thus, stations can save much power. The network bandwidth can be efficiently utilized as well. Simulation results show that PEM outperforms than existing protocols, such as DCF, DPSM, and DCS. PEM not only reduces power consumption, but also increases the network throughput.

### 3.12. MT-MAC (Multi-hop TDMA Energy-efficient Sleeping MAC Protocol)

In order to improve the performance of energy efficiency, throughput and delay in WMSN, the authors [34] designed a Multi-hop TDMA Energy-efficient Sleeping MAC (MT-MAC) protocol. The main idea of MT-MAC is to divide one frame into many slots for sensor nodes in WMSN to forward data packets, with TDMA scheduling method. At first, each node tries to get routing information from its routing layer while constructing the neighbor list through flooding algorithm. Then, after collecting the necessary information, each node transfers both the routing information and neighbor information to the mesh router. And the mesh router computes the slot allocation table called the schedule. At last, the mesh router broadcasts both the schedule and its own Synchronous clock. So every node not only keeps synchronized with the mesh router, but also decides when to forward data according to the schedule. As collision and hidden terminal problems are considered in the scheduling algorithm, so the probability of collision in the network will be significantly reduced, and the network throughput will be improved. As the simulation result showed, compared with SMAC protocol, MT-MAC not only saves 25% energy consumption of the network, but also decreases 45% latency of the whole network.

**Summary**

The discussed MAC protocols are summarized and classified in the table below.

| Protocol | Network | Topology | Channel/ Synchronization | Contribution |
|---|---|---|---|---|
| PSM [22] | 802.11 | Flat | 1 / Yes | Active/sleep period |
| APSD [23] | 802.11e | Flat | 1 / Yes | Active/sleep schedules |
| PSMP [24] | 802.11n | Flat | 2 / Yes | Active/sleep schedules with reducing overheads |
| NAV [25] | 802.11 | Flat | 1 / Yes | Active/sleep with blockage of transmission when channel is reserved |
| MAN [26] | 802.11 | Flat | 1 / Yes | Map of active/sleep period |

| PAMAS [27] | Ad hoc | Flat | 2 / No | Wake up radio scheme |
| --- | --- | --- | --- | --- |
| SOFA [28] | 802.11 | Flat | 1 / NO | A round-robin scheduling and sleep/wakeup |
| PRCW [29] | 802.11 e | Flat | 1 / NO | Admission control algorithm |
| LEACH [30] | 802.15.4 | Clustered | 1 / Yes | Low energy clustering |
| S-MAC [31] | 802.15.4 | Flat | 1 / Yes | Active/sleep schedules |
| Virtualization of NICs [32] | 802.11 | Flat | 1 / Yes | Virtualization of network interface cards |
| PEM [33] | 802.11 | Flat | 1 / No | Power control and power saving mechanisms |
| MT-MAC [34] | WMSN | Clustered | 1 / Yes | TDMA Energy-efficient Sleeping MAC |

The Data Link Layer is very important to ensure energy conservation for Wireless Mesh Networks. Many researchers are focused on specific approaches as we have seen previously. The most of researches used power save mode mechanism based on 802.11 standard or the TDMA scheduling method to improve and better energy saving.

## 4. ENERGY CONSERVATION IN PHYSICAL LAYER

Physical Layer (PHY) implements a network communication hardware, which transmits and receives messages one bit or symbol at a time. The researches that involve the Physical Layer of WMNs are pointed out in this Section.

### 4.1. The LM-SPT (Local Minimum Shortest Path-Tree)

This work [35] addressed the challenge of power efficiency by presenting a localized distributed power-efficient topology control algorithm for application of WMNs in rural areas. The main idea in this approach was to construct an overlay graph topology of the WMN with such desirable features as reduced physical node degree, increased throughput, increased network lifetime and maintenance of connectivity by varying the transmission power at each node. The main contribution of this work was the ability of the algorithm to balance energy efficiency and throughput in WMNs and without loss of connectivity. The concept of this approach was based on information of the local neighborhood that is confined to one hop from the logical visible neighborhood for calculating the minimum power transmission. In this approach Aron et al.[9] proposed to calculate minimum transmission power and require only an optimal value to maintain an optimal connectivity and decrease the interference and collision which save energy and achieve a good throughput.

### 4.2. CNN (Critical Number of Neighbors)

Santi [36] used CNN refers to the minimum number of neighbors that should be maintained by each node in the network to be asymptotically connected. This approach to maintain connectivity is adopted for the use in the proposed scheme because only knowledge of the network size is required to determine the CNN. This information can be easily obtained from a proactive routing protocol such as OLSR. The CNN may also result in heterogeneous transceiver power outputs, potentially maximizing power savings and interference gains. The CNN is also less affected by the distribution and position of the network nodes and there is no need to assume a GPS enabled router. It's also increases gradually with network size and is thus able to tolerate delays in the propagation of topology updates and network size (if a proactive routing protocol is used). Thus, maintaining connectivity via a CNN will reduce human intervention (if a proactive routing protocol is employed). A similar idea to Aron et al. [35] approach, Santi [36] proposed to calculate minimum number of neighbors that should be maintained to be connected to conserve energy.

### 4.3. Minimum-energy topology

Aron et al. [37] considered the problem of topology control in a hybrid WMN of heterogeneous and presented a localized distributed topology control which pointed to calculate the optimal transmission power to maintain connectivity and reduce the transmission power to cover only nearest neighbors with saving energy and extending the lifetimes of the networks. In this approach, the main objective was to generate a minimum-energy topology G graph. Taking an arbitrary node $u \in V$ in the network G, a three phased topology control algorithm those run in each node. First, the node $u$ broadcasts a "Hi" message using its full power, $P^{max}_u$. The nodes that receive the "Hi" message form the set of accessible neighborhood of node $u$. The "Hi" message contains the id of u, the location information of $u$, (x (u), y (u)) and the value of the $P^{max}_u$. After there exist a weighted directed graph topology G, $u \in V$. Node $u$ has knowledge of the edge weights and path weights, where path weight of a directed path. Finally, the node $u$ determines its own transmission power and the power on the accessible edges of all the nodes in the accessible neighborhood. Node $u$ takes as its power, the largest one-hop edge weight among the edges obtained in the minimum-energy local topology view G, after node u adopt its minimum-energy level; it propagates this minimum power value to the other neighbors in the accessible neighborhood with the current Transmission power. Aron et al. [37] developed a minimum-energy distributed topology control that ensures a reduction in the amount of energy consumed per node during transmissions and without loss of connectivity with increasing in the network lifetime.

### 4.4. PlainTC

Mudali et al. [38] investigate the feasibility of power control in a popular WMN backbone device and design and evaluate an autonomous, light-weight TC scheme called PlainTC. Two main approaches may be used in this regard, either maintaining the Critical Transmission Range

(CTR) or the Critical Neighbor Number (CNN). In this work, Mudali et al. presented a TC scheme for a WMN backbone comprising of commercially available Linksys WRT54GL routers (which are popular WMN backbone devices). The proposed scheme is designed to maintain network connectivity by relying on data collected by a proactive routing protocol. Three types of information can be collected and used as the basis of a TC scheme: location information, direction information and neighbor information. The Linksys WRT54GL device contains neither a GPS nor the native ability to determine the relative direction of incoming and outgoing transmissions. The device does however possess the ability to collect low-quality, neighbor-based information.by inspecting the routing table built by the proactive routing protocol being employed. Determining the logical node degree is easier because the number of HELLO messages received from unique sources can be determined if a reactive routing protocol is employed. If a proactive routing protocol is employed, then the routing table can be inspected for the number of one-hop (or n-hop) neighbors. The approach proposed by Mudali et al. [38], is designed for a specific device (Linksys WRT54GL) and it contains a GPS to define the direction. Unfortunately, generally the devices are not specific and they are not equipped with a GPS generally.

### 4.5. Virtual WLAN

Coskun et al. [39], proposed the switching scheme which aims to powering on the minimum number of devices or the combination of devices that consume the least energy that can jointly provide full coverage and enough capacity. This approach corresponds to having a minimum set of devices that provide coverage and an additional set of devices that are powered on to provide additional capacity when needed. By consolidating hardware, some hardware can be put in low-power mode and energy consumption can be reduced, saving the difference in energy consumption per low-power node when compared to an active node and adding the amount of energy consumed by hosting more networking processes just on fewer network nodes. Depending on the type of device, different amounts can be saved. Using virtual 802.11 interfaces to connect to multiple networks simultaneously, instead of using multiple network interface cards, enables savings in energy costs, minimizes the physical space, and provides the capability to build large and small wireless mesh networks.

### 4.6. Sleeping and Rate-Adaptation

Nedevschi et al. [40] presented the design and evaluation of two forms of power management schemes that reduce the energy consumption of networks. The first puts network interfaces to sleep during short idle periods. To make this effective they introduce small amounts of buffering, much as 802.11 APs (access points) do for sleeping clients; this collects packets into small bursts and thereby creates gaps long enough to profitably sleep. Potential concerns are that buffering will add too much delay across the network and that bursts will exacerbate loss. Their algorithms arrange for routers and switches to sleep in a manner that ensures the buffering delay penalty is paid only once (not per link) and that routers clear bursts so as to not amplify loss noticeably. The result is a novel scheme that differs from 802.11 schemes in that all network elements are able to sleep when not utilized yet added delay is bounded. The second approach adapts the rate of individual links based on the utilization and queuing delay of the link.

### 4.7. Connectivity strategies

Mudali et al. proposed [41]; various connectivity strategies based on the Critical Number of Neighbors approach are evaluated via simulation to determine the relationship between transceiver power savings and the network lifetime. The evaluation indicates that the selected connectivity strategies are able to produce cumulative transceiver power savings. The extent of the power savings produced by individual backbone nodes is largely dependent upon the location of the node relative to the (imaginary) center of the backbone network. The evaluation also suggests that cumulative transceiver power savings do not automatically translate into corresponding extensions of network lifetime.

### 4.8. Greening of the Internet

Gupta and Singh [42] identify the problem of excessive energy consumption in the Internet and propose sleeping as the approach to save energy. They thought that sleeping was an appropriate way to maximize internet's energy conservation. However, in order to implement algorithms for sleeping, firstly a redesigning of the network equipment's hardware was done to allow software-enabled sleeping, secondly the routing protocols was modified so as to adapt energy consumption and allow its loading via aggregation and sleeping, thirdly, for more options of route selection allowing aggregation and sleeping, the Internet topology was amended as well, and, finally studying the impact of sleeping on protocols such as TCP with an eye on changing the protocol so as to adapt to the presence of sleeping nodes. It appeared that sleeping was indeed a feasible strategy but it will require some changes to the current protocol specifications.

Further, in order to maximize the amount of energy conservation, they note that some modifications to the Internet architecture may be needed (particularly adding more links to allow packet aggregation along fewer routes).

### 4.9. Power Control and Rate Adaptation

In [43], authors study the scheduling optimization problem in wireless MESH networks assuming a time division multiple access (TDMA) scheme, a dynamic power control able to vary emitted power slot-by-slot, and a rate adaptation mechanism that sets transmission rates according to the SINR. Traffic quality constraints were expressed in terms of minimum required bandwidth. Since the time frame defined by the TDMA scheme is fixed, the bandwidth requirement can be translated into the number of information units (packets) that must be transmitted on each link per frame. Moreover, according to a discrete set of possible transmission rates, the number of packets that can be transmitted per time slot depends on the SINR at receivers. In order to get more insights on the characteristics of the problem and the effect of different control mechanisms, they considered three different versions of the problem with increasing complexity. In the first one they assume fixed power and rate, in the second one variable power and fixed rate, and finally in the third one variable power and rate. Given a number of available slots, the goal is to provide an assignment of time slots to links such that bandwidth constraints are satisfied and the number of available time slots is not exceeded.

### 4.10. Energy savings in Wireless Mesh Networks in a time-variable context

In [5], the authors proposed a novel approach for the dynamic energy management of WMNs to minimize the energy in a time varying context by selecting dynamically a subset of mesh BSs to switch on considering coverage issues of the service area, traffic routing, as well as capacity limitations both on the access segment and the wireless backhaul links. To reach the objective, they provided an optimization framework based on mathematical programming that considers traffic demands for a set of time intervals and manages the energy consumption of the network with the goal of making it proportional to the load. The contributions provided by the authors are: Firstly, they took account of the access segment and also the wireless backhaul of wireless access networks. Secondly, they combined together the issue of wireless coverage, for the access segment, and the routing, for the backhaul network, and optimize them jointly. Thirdly, the authors explicitly include traffic variations over a set of time intervals and show how it is possible to have energy consumption following these variations. Finally, they provide a rigorous mathematical modelling of the energy minimization problem based on Mixed Integer Linear Programming (MILP), and solve it to the optimum.

**Summary**

Topology control algorithms have largely been proven to be one way of achieving energy efficiency in MWNs. Several contributions have been tailored towards studying power control problems in energy-constrained conventional IEEE 802.11 wireless network standards; little attention has been drawn to the power control problems in WMNs. Control of topology and control of power transmission are the most widely used, simple but very effective for saving energy,

## 5. CONCLUSIONS

Various existing energy conservation methods proposed by different studies but the energy is still a challenging issue for wireless mesh network because of its significant environmental footprint. In this paper, we have summarized some research results which have been presented in the literature on energy saving methods in wireless mesh networks. Although many of these energy saving techniques look promising, there are still many challenges that need to be solved. Therefore, further research is necessary for handling these kinds of situations. On the basis of these studies have been previously conducted by different researchers we reach at the obvious conclusion that if we want to conserve energy in WMN, the most effective way is to combine the most effective solutions in the three lower layers but it still a challenge.

## REFERENCES

[1] I.F. Akyildiz and X. Wang, (2005) "A survey on wireless mesh networks". IEEE communications Magazine 43(9), s23–s30.

[2] I.F. Akyildiz, X. Wang and W. Wang, (2005) "Wireless mesh networks: a survey". Computer Networks Journal (Elsevier) 47(4), pp 445–487.

[3] The Wi-Fi Alliance. http://www.wi-fi.org/.

[4] The WiMAX Forum. http://www.wimaxforum.org/home.

[5] A. Capone, F.Malandra, B. Sansò, (2012) "Energy Savings in Wireless Mesh Networks in a Time-Variable Context". Mobile Networks and Applications, Volume 17, Issue 2, pp 298-311.

[6] S. Giuannoulis, C. Antonopoulos, E. Topalis, and S. Koubias, (2005) "ZRP versus DSR and TORA: A comprehensive survey on ZRP performance".10th IEEE Conference on Emerging Technology Automation, 8 pp. – 1024.

[7] Michael Bahr; (2006) "Proposed routing for IEEE 802.11s WLAN mesh networks". WICON '06 Proceedings of the 2nd annual international workshop on Wireless internet, Article No.5.

[8] M. Cardei, M.X. Cheng, X. Cheng, Du D.-Z., (2002) "Connected Domination in Ad Hoc Wireless Networks", Proceedings of the Sixth International Conference on Computer Science and Informatics (CSI).

[9] B.Chen, K. Jamieson, H. Balakrishnan and R. Morris, (2002) "Span: An Energy-Efficient Coordination Algorithm for Topology Maintenance in Ad Hoc Wireless Networks", Kluwer Academic Publishers. Manufactured in The Netherlands, 2002.

[10] Y. Xu, J. Heidemann and D. Estrin, (2001), " Geography-informed energy conservation for ad hoc routing", Proceedings of the Seventh Annual ACM/IEEE International Conference on Mobile Computing and Networking (MobiCom), 2001.

[11] B. Bouyedou , M.FEHAM, F. Didi, H. Labiod,(2009) "Improvement of DSR Performances in Mobile Ad hoc Networks with Trade-off between Energy Efficiency and Path Shortness", IEEE

[12] T. Sheppard, "A channel access scheme for large dense packet radio networks", Proceedings of the ACM SIGCOMM, 1996.

[13] J. Chang and L. Tassiulas, "Energy conserving routing in wireless ad hoc Networks", Proceedings of the Fifth Annual ACM/IEEE International Conference on Mobile Computing and Network (MobiCom), Dallas,TX (August 1998).

[14] L. Lin, N.B. Shroff, and R. Srikant. "Asymptotically Optimal Power- Aware Routing for Multihop Wireless Networks with Renewable Energy Sources". INFOCOM 2005.

[15] B. Awerbuch, D. Holmer, and H. Rubens, "The Pulse Protocol: Energy Efficient Infrastructure Access", IEEE INFOCOM, 2004.

[16] Iannaccone, Jardosh, Papagiannaki and Vinnakota, (2007) "Towards an Energy-Star WLAN infrastructure", Mobile Computing Systems and Applications, Hotmobile., Eighth IEEE Workshop., pp. 85-90, 2007.

[17] Chen, Chin, Yang and Yin, "Power-Efficient Access-Point Selection for Indoor Location Estimation", IEEE Transactions and Data Engineering, Vol. 18, No.7, pp. 877-888, DOI:

[18] J. Restrepo, C. Gruber, and C. Machuca, (2009), "Energy profile aware routing," in Communications Workshops, 2009. IEEE International Conference on, pp. 1 –5.

[19] A. Oliva, A. Banchs and P. Serrano, (2012), "Throughput and energy-aware routing for 802.11 based mesh networks", Comput. Commun. http://dx.doi.org/10.1016/j.comcom.2012.04.004.

[20] Amokrane Ahmed, Langar Rami, Boutaba Raouf and Pujolle Guy; (2012), "A Green Framework for Energy Efficient Management in TDMA-based Wireless Mesh Networks"; IEEE/ACM CNSM 2012 , Las Vegas, USA.

[21] C. BEMMOUSSAT, F. DIDI, M. FEHAM, (2012) "EFFICIENT ROUTING PROTOCOL TO SUPPORT QOS IN WIRELESS MESH NETWORK", International Journal of Wireless & Mobile Networks (IJWMN) Vol. 4, No. 5, October 2012.

[22] IEEE 802.11, Wireless LAN Medium Access Control (MAC) and Physical Layer (PHY) Specifications, 1999.

[23] IEEE Stds. Dept.. "Part 11: Wireless Medium Access Control (MAC) and Physical Layer (PHY) Specifications: Medium Access Control (MAC) Quality of Service (QoS) Enhancements", IEEE Press, 2005.

[24] IEEE Stds. Dept., "Part 11: Wireless LAN Medium Access Control (MAC) and Physical Layer (PHY) specifications: Amendment 4: Enhancements for Higher Throughput", IEEE P802.11n/D3.00. IEEE Press, 2007.

[25] F. Zhang et al., "Power Saving Access Points for IEEE 802.11 Wireless Network Infrastructure" , IEEE Wireless Commun. and Networking Conf. 2004, March 2004.

[26] Y. Li, T. D. Todd, and D. Zhao. "Access Point Power Saving in Solar/Battery Powered IEEE 802.11 ESS Mesh Networks", 2nd Int'l. Conf. Quality of Service in Heterogeneous Wired/Wireless Networks, Augst 2005.

[27] S.Singh and C.S.Raghavendra, "Power aware multi-access protocol with signaling for ad hoc networks", ACM Computer Communication Review, Vol. 28 No. 3, pp. 5-26, July 1998.

[28] Z. Zeng, Y. Gao, and P. R. Kumar, SOFA: A Sleep-Optimal Fair-Attention scheduler for the Power-Saving Mode of WLANs, 31st International Conference on Distributed Computing Systems, 2011 IEEE.

[29] F. Didi, H. Labiod, G. Pujolle, M. Feham, "Physical Rate and Contention Window based admission control (PRCW) for 802.11 WLANs", IEEE Symposium on Computers and Communications, Riccione, Italy, August, 2010.

[30]   W.R. Heinzelman, A. Chandrakasan, and H. Balakrishnan. "Energy-Efficient Communication Protocol for Wireless Micro Sensor Networks". In IEEE Proceedings of the Hawaii international Conference on System Sciences (HICSS '00), 2002.

[31]   W. Ye, J. Heidemann, D. Estrin, "An Energy- Efficient MAC Protocol for Wireless Sensor Networks", IEEE INFOCOM 2002.

[32]   Y. Al-Hazmi and Hermann, "Virtualization of 802.11 Interfaces for Wireless Mesh Networks", Wireless On-Demand Network Systems and Services (WONS), 2011 Eighth International Conference on 26-28 Jan 2011.

[33]   K. Shih, C. Chang, C. Min Chou, S. Chen, "A Power Saving MAC Protocol by Increasing Spatial Reuse for IEEE 802.11 Ad Hoc WLANs", AINA '05 Proceedings of the 19th International Conference on Advanced Information Networking and Applications - Volume 1 Pages 420-425

[34]   Q. Fan, J. Fan1, J. Li, and Xiaofang Wang; (2012), "A Multi-hop Energy-efficient Sleeping MAC Protocol based on TDMA Scheduling for Wireless Mesh Sensor Networks", JOURNAL OF NETWORKS, VOL. 7, NO. 9, SEPTEMBER 2012

[35]   F.O. Aron, A. Kurien and Y. Hamam, " Topology Control Algorithm for Effective Power Efficiency and Throughput for Wireless Mesh Networks", Third International Conf. on Broadband Communications, Information Technology & Biomedical Applications, 2008.

[36]   P.Santi, "Topology Control in Wireless Ad Hoc and Sensor Networks", Wiley: Chichester, 2005.

[37]   F. O. Aron, T. O. Olwal, A. Kurien, and M. O. Odhiambo, "A Distributed Topology Control Algorithm to Conserve Energy in Heterogeneous Wireless Mesh Networks", World Academy of Science, Engineering and Technology 40, 2008.

[38]   P. Mudali, T.C. Nyandeni, N. Ntlatlapay, and M.O. Adigun, "Design and Implementation of a Topology Control Scheme for Wireless Mesh Networks", IEEE AFRICON 2009.

[39]   H. Coskun, I. Schieferdecker and Y. Al-Hazmi, "Virtual WLAN: Going beyond Virtual Access Points", ECEASST, vol. 17, March 2009.

[40]   S. Nedevschi, L. Popa, G. Iannaccone, "Reducing Network Energy Consumption via Sleeping and Rate-Adaptation", Proceedings of the 5th USENIX Symposium on Networked Systems Design and Implementation (2008), pp. 323-336.

[41]   P. Mudali, T.C. Nyandeni, N. Ntlatlapay, and M.O. Adigun, "Design and Implementation of a Topology Control Scheme for Wireless Mesh Networks", IEEE AFRICON 2009.

[42]   M. Gupta and S. Singh, Greening of the Internet, SIGCOMM'03, August 25–29, 2003, Karlsruhe, Germany. Copyright 2003 ACM.

[43]   A. Capone and Giuliana, "Scheduling Optimization in Wireless MESH Networks with Power control and Rate Adaptation", Sensor and Ad Hoc Communications and Networks, 2006. SECON '06. (2006), 3rd Annual IEEE Communications Society.